\documentclass[fleqn,12pt,twoside,epsfig]{article}
\usepackage{espcrc1}

\usepackage{graphicx}
\usepackage[figuresright]{rotating}

\newcommand{\AmS}{{\protect\the\textfont2
  A\kern-.1667em\lower.5ex\hbox{M}\kern-.125emS}}

\title{Three-body model for the complete fusion of a two-cluster
 composite projectile with a heavy target}

\author{M.S. Hussein\address[USP]{Instituto de F\'isica, Universidade de S\~ao Paulo, 
        C.P. 66318, 05315-930, S\~ao Paulo, SP, Brazil}, 
        B.V. Carlson\address[CTA]{Departamento de F\'isica, Instituto
          Tecnol\'ogico da Aeron\'autica, CTA, 12228-900, S\~ao Jos\'e
          dos Campos, SP, Brazil },
        T. Frederico\addressmark[CTA] and
        T. Tarutina\addressmark[USP]}
       
\begin{document}

\maketitle

\begin{abstract}
The purpose of this work is to show that within a three-body
description, the complete fusion process and the target excitation by
the projectile can be taken into account by introducing a three-body
optical potential in the formalism. We give a schematic description of
such a potential and points to ways of testing the validity of the CDCC.

\end{abstract}

\section{Introduction}
The fusion of exotic nuclei with heavy targets has recently been the
subject of intense experimental and theoretical investigation
\cite{Ber93}. 

Recently a formalism has been introduced \cite{Can98} in the study
of the incomplete fusion process resulting in a better understanding
of its physical characteristics as well as giving the possibility of
unifying various, and in some cases somewhat conflicting approaches.
In the model, consisting of a target $A$ and a projectile $a$ composed
of two particles $b$ and $x$, the process is described by the optical
potentials $U_{xA}$ and $U_{bA}$ of the systems $(xA)$ and $(bA)$, 
respectively, as well as the hermitian potential $V_{xb}$ which binds
the structure-less particles $b$ and $x$ to form the projectile $a$. The
incomplete fusion process appears directly connected to the absorptive
 (imaginary) parts of $U_{xA}$ and $U_{bA}$, denoted by $W_{xA}$ and $W_{bA}$,
 respectively. In this framework, the total absorption cross section
 is given by
\begin{equation}
\sigma_{abs}\propto\langle\phi^{(+)}_x|W_{xA}|\phi^{(+)}_x\rangle + 
\langle\phi^{(+)}_b|W_{bA}|\phi^{(+)}_b\rangle,
\label{first}
\end{equation}
where each term of Eq.(\ref{first}) represents the integration of the
incomplete fusion cross section of observing the spectator particles $x$ 
and $b$. The wave functions $|\phi^{(+)}\rangle$ are source functions \cite{Can98}.
The complete fusion cross section is obtained from 
$\langle\phi^{(+)}|W_{aA}|\phi^{(+)}\rangle$ after subtracting
(\ref{first})
and the elastic break-up contribution.

To be consistent, however, one should only consider $U_{xA}$ and
$U_{bA}$, within a genuine three-body model \cite{Hus90}. The
three-body model, however, does not account for the complete fusion of
the particle $a$ nor for the excitation of the target $A$ by the
projectile. These processes are missing in the three-body model due,
as we shall see, to the use of two-body optical potentials.

When calculating complete fusion, we have difficulty in evaluating
consistently incomplete fusion (or inclusive breakup). The schematic
figures below show the two processes.

\begin{figure}
\includegraphics[scale=0.5]{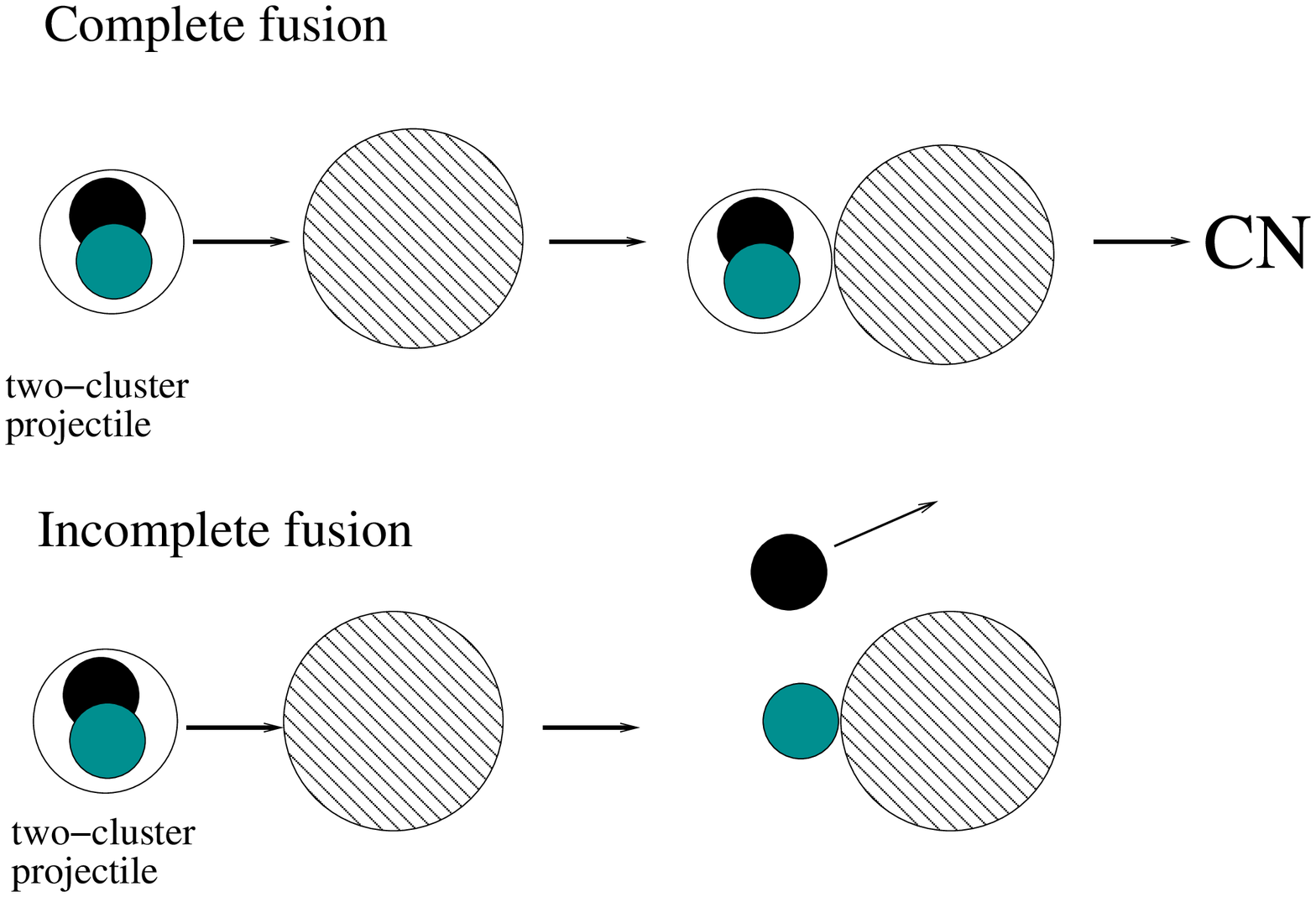}
\caption{}
\end{figure}

On the other hand, when calculating incomplete fusion, one gets an
underestimation of total fusion (= complete + incomplete fusion),
because complete fusion is not included (difficult). Accordingly,
calculation of fusion of a two-cluster projectile with a target
nucleus is difficult.

\section{Theories of Incomplete and Complete fusion}
In the eighties, Austern {\it et al} \cite{Aus87}, Ichimura \cite{Ich90},
Hussein and McVoy \cite{Hus85}, Udagawa and Tamura \cite{Uda86} and 
others worked hard to formulate
a practical and consistent three-body theory of inclusive breakup
reactions (incomplete fusion). They did not address the question
of complete fusion. Here we give some thoughts concerning this
issue.

In the spectator model, cluster $b$ in the projectile $a=b+x$,
only scatters optically from the target.
It is proven by Austern et al.\cite{Aus87} that the incomplete fusion of the
participant particle $x$ is given by the energy and angle
integrals of the spectator particle $b$'s spectrum

\begin{equation}
\sigma_{IF}=\int d\Omega_b\frac{d^2\sigma}{d\Omega_bdE_b},
\end{equation}
where
\begin{equation}
\frac{d^2\sigma}{d\Omega_bdE_b}=\rho(E_b)\int d{\vec r_x}
\langle\Psi_{aA}^{(+)}|\chi_b^{(-)})({\vec r_x})W_{xA}({\vec
r_x})(\chi_b^{(-)}|\Psi_{aA}^{(+)}\rangle({\vec r_x}),
\end{equation}
$|\Psi_{aA}^{(+)}\rangle$ is the full three-body wave function.

In the limit of $\Psi_{aA}^{(+)}\longrightarrow\chi_{aA}^{(+)}$,
one gets the Hussein-McVoy theory \cite{Hus85} whose no-distortion limit (of
$b$) is just the old Serber model. M.Ichimura clarified the
connections among the different theories \cite{Ich90}. Notice that the
source functions $|\phi^{(+)}_x\rangle$ and $|\phi^{(+)}_b\rangle$ of
Eq.(\ref{first}) are just $\langle\chi_b^{(-)}|\chi_{aA}^{(+)}\rangle$ and
$\langle\chi_x^{(-)}|\chi_{aA}^{(+)}\rangle$.

In the Glauber limit,
\begin{equation}
\sigma_{IF}^{(x)}=\frac{\pi}{k^2}\sum_{l_x}(2l_x+1)\langle
T_x(l_x)(1-T_b(l_b))\rangle
\end{equation}

Though not apparent, Yabana {\it et al.} \cite{Yab03} hinges on the above (no
$b$-target interaction) and thus they underestimate the total fusion.

To obtain complete fusion, one "guesses" the result using
unitarity. This is not consistent three-body model of complete
fusion.

On the other hand, to account for CF using Faddeev equations with non-hermitian
coupling hamiltonians, one is bound to introduce an effective
three-body optical potential. We turn to this question in what follows.

Within the Feshbach projection operator framework, the part of the
optical potential that arises from the $xA$ coupling is:
\begin{equation}
\hat U_{xA} =  V_{xA;A+x}^+G_{A+x}V_{xA;A+x}
\end{equation}
\vskip 1 cm
\includegraphics[scale=0.6]{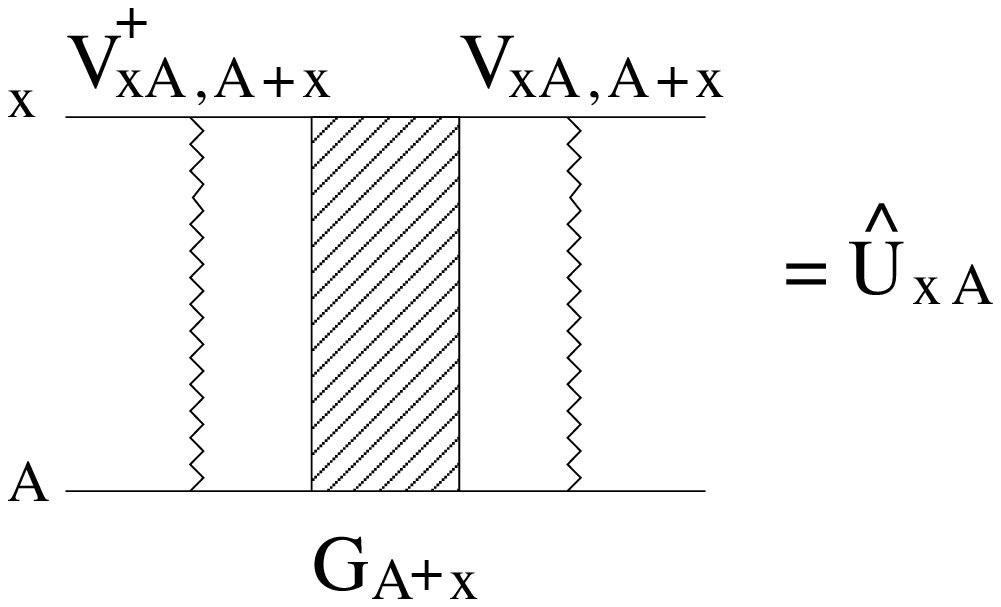}\\

There is a similar potential for $b$ (x taken as spectator)
\begin{equation}
\hat U_{bA} =  V_{bA;A+b}^+G_{A+b}V_{bA;A+b}
\end{equation}

Accordingly a typical term in the $a(=x+b)+A$ T-matrix:
\vskip 1 cm
\includegraphics[scale=0.6]{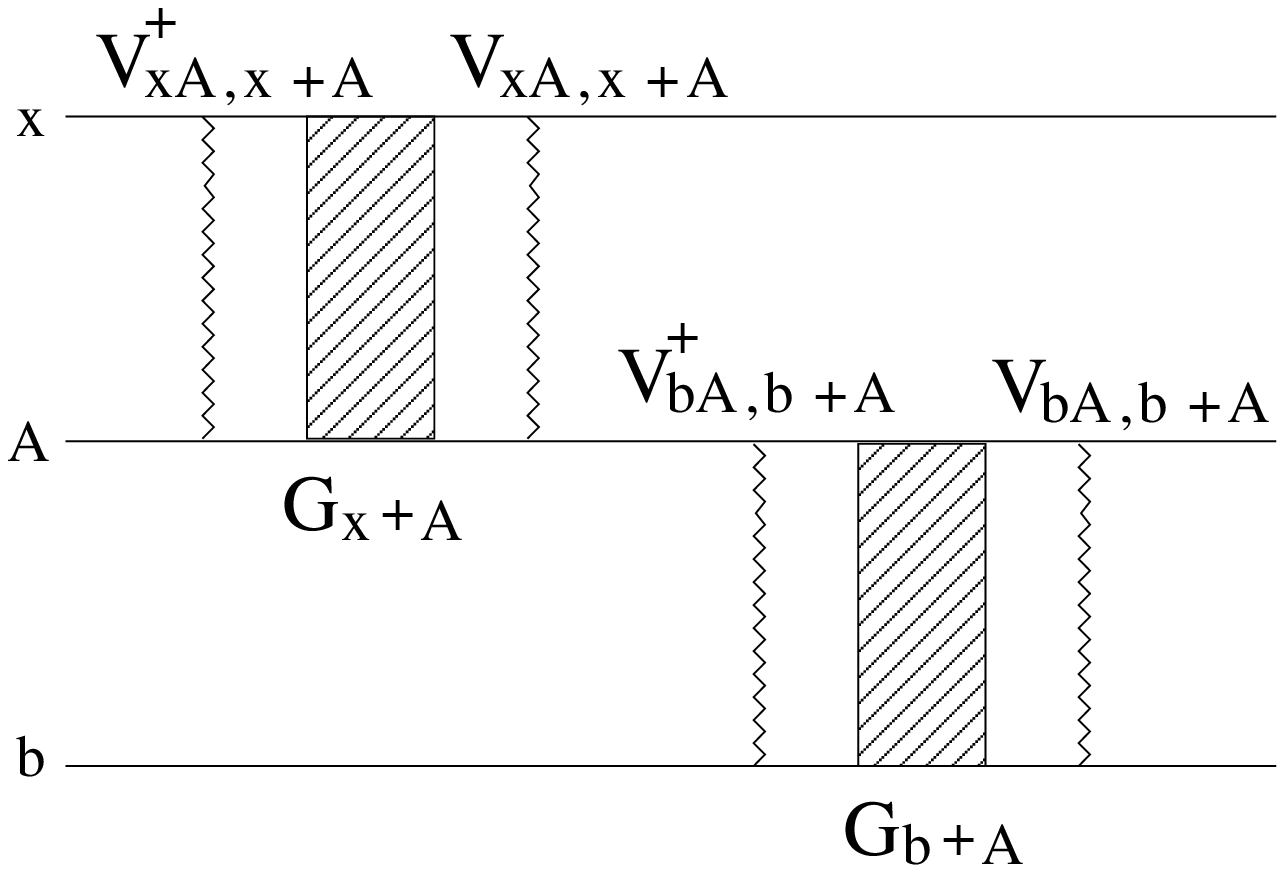}\\

Thus during the intermediate $A+x$ or $A+b$ propagation, $x$ and
$b$ do not interfere with each other; $x$ and $b$ are never
simultaneously absorbed by target: no complete fusion.

To include CF, the spectator particle (be it $b$ or $x$), must be
allowed to interact with the intermediate composite system.

A possible candidate for three-body optical potential is:
\begin{equation}
U_{Axb}=V^+_{xA,A+x}G_{A+x,b}V^+_{b,A+x;A+x+b}G_{A+x+b}
V_{x,A+b;A+x+b}G_{A+b,x}V_{bA,A+b}
\end{equation}
\vskip 1 cm
\includegraphics[scale=0.6]{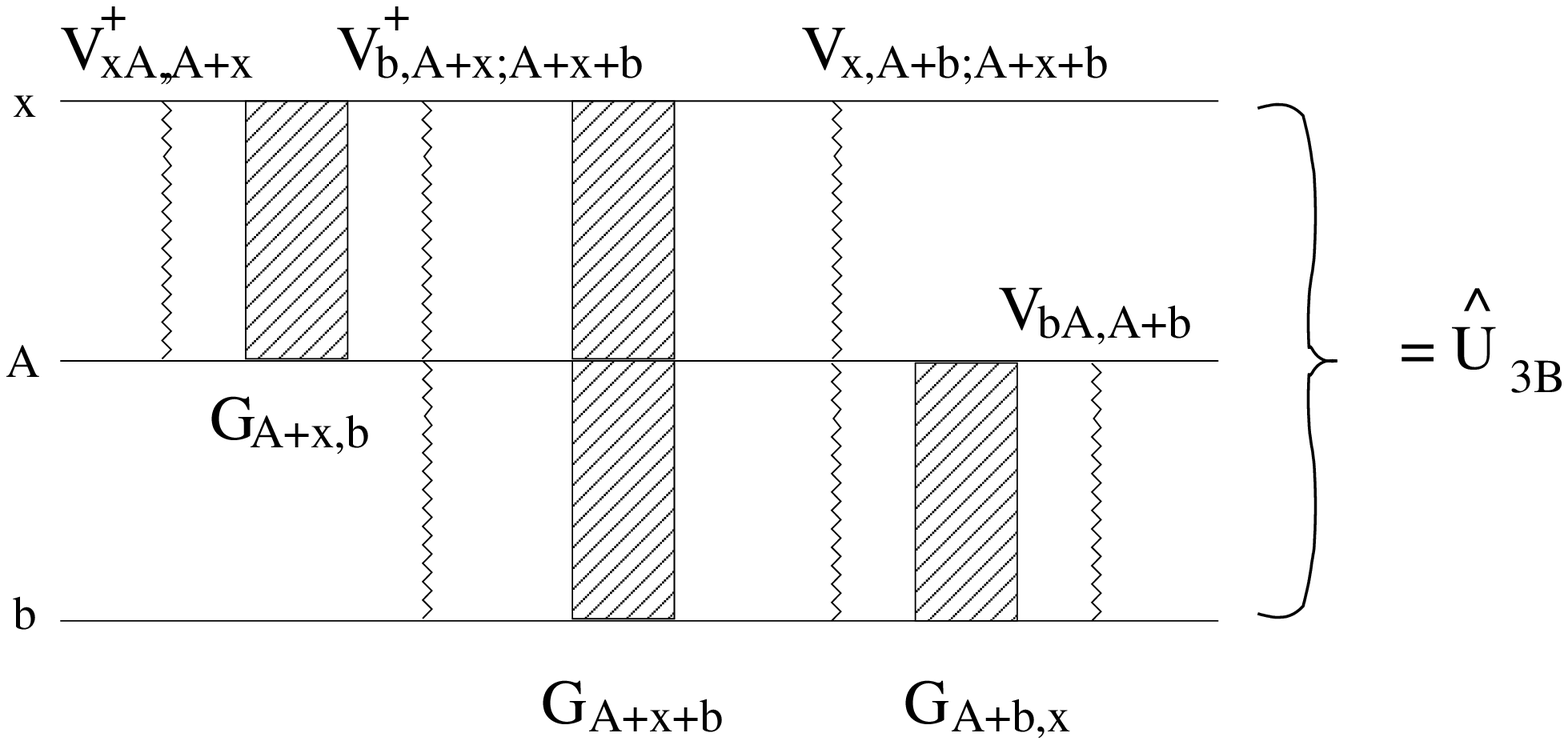}\\

Thus, with $W_{xA}$, $W_{bA}$ and $W_{3B}$, being the imaginary
potentials of the $x+A$, $b+A$ and $x+b+A$ systems respectively,
we have for incomplete fusion:
\begin{equation}
\sigma_{IF}\propto\int\langle\Psi^{(+)}|\chi_b^{(-)})({\vec
r_x})W_{xA} ({\vec r_x})(\chi_b^{(-)}|\Psi^{(+)}\rangle d{\vec r_x}
\label{sigmaIF_1}
\end{equation}
or
\begin{equation}
\sigma_{IF}\propto\int\langle\Psi^{(+)}|\chi_x^{(-)})({\vec
r_b})W_{bA} ({\vec r_b})(\chi_x^{(-)}|\Psi^{(+)}\rangle d{\vec r_b}
\label{sigmaIF_2}
\end{equation}
and the complete fusion
\begin{equation}
\sigma_{CF}\propto\langle\Psi^{(+)}|W_{3B}|\Psi^{(+)}\rangle.
\label{sigmaIF_3}
\end{equation}
Eqs.(\ref{sigmaIF_1}) and (\ref{sigmaIF_2}) have been analyzed in
Ref.\cite{Can98}. It remains as a challenge the calculation of
Eq.(\ref{sigmaIF_3}) \cite{Car03}.

\section{Conclusions}

Within a three-body, Faddeev, description of the fusion reaction
one needs three-body optical potential (simplified version of $\hat
U_{3B}$) to account for  $\sigma_{CF}$. Such a description will allow assessment of
the CDCC model which consistently fails to account for the incomplete
fusion \cite{Dia02}.

This work is supported by FAPESP and CNPq.

\end{document}